\documentstyle[preprint,aps]{revtex}
\tightenlines

\begin{document}
\bibliographystyle{unsrt}
\input{psfig}

\date{\today}

\title{Soliton mediated protein folding}
\author{S. Caspi and  E. Ben-Jacob}

\address{School of Physics and Astronomy, Tel Aviv University,
69978 Tel Aviv, Israel.}

\date{\today}
\maketitle

\begin{abstract}
A novel approach to protein folding dynamics is presented. 
We suggest that folding of protein may be mediated via interaction with solitons
which propagate along the molecular chain. A simple toy model is presented in
which a Sine-Gordon field interact with another field $\phi$ corresponding to the 
conformation angles of the protein. We demonstrate how a soliton carries this
field over energy barriers and consequently enhances the rate of the folding
process. The soliton compensate for its energy loss by pumping the energy gained 
by the folded field. This scenario does not change significantly even in the
presence of dissipation and imposed disorder.
\end{abstract}
\vspace*{.2in}

The possibility for existence of solitons in bio-molecules has been widely studied
since the work by Davidov which described solitons in proteins as means to 
transfer energy and charge \cite{Davidov1}. It is evident today that solitonic 
excitation should
be included in order to describe properly the biophysics of proteins and DNA
molecules ({\it e.g.\/} see \cite{Solitons_in_Proteins1}, \cite{Open_state_soliton_in_DNA1}, 
\cite{Charge_soliton_in_DNA1}), though their importance to the biological activity is still 
controversial. 

In this paper we suggest that solitons may take part in the process of protein
folding. The way in which a protein, a linear chain of amino acids, folds into
a well defined, three dimensional native structure, is still far from being
understood, although considerable effort is devoted to the subject. Since
many biological proteins fold correctly without assistance, it is clear that
all the information about their native state exists already in the sequence of
amino acids \cite{Biochemistry}. The folding dynamics may be viewed as navigation
process in the energy landscape of the protein conformational space. 
It is believed that the native structure 
corresponds to a small region in that space located in the vicinity of 
the ground state. Identifying this ground
state is highly non trivial, but this is not the entire story. The ground state
should be accessible dynamically within time scales not larger than minutes. As 
pointed out by Levinthal, stochastic search for the ground state over a random
landscape might take cosmological time. From this fact one may draw the conclusion
that the landscape of natural proteins is not random, but, rather, directed 
towards the ground state \cite{Funnel_fold1}. Conventionally, overcoming the
energetic barriers is ascribed to a stochastic and uncorrelated thermal activation.
Here we shall explore another option. Surly, random thermal processes have important 
role in folding dynamics, but it may be that other processes, correlated and 
deterministic, dictates the folding pathway.
To illustrates what is meant by this, suppose that energy gained at a point
where the molecule already reached the lower energy
state, could propagate efficiently to another point along the molecule
and induce consequent fold there. This may go on further, as more energy is
now released. In this way, large sections of a protein can fold very fast,
and moreover, the folding process can be orchestrated deterministically. 
Solitons provide a mechanism for such stable, non dispersive energy transfer.
We shall refer to a folding process induced by the propagation of a solitonic 
wave along the molecular chain as a Soliton Mediated Folding (SMF).

In order to explore the possibilities of soliton interaction with 
local conformation angles of the molecular chain, we
introduce a toy-model inspired by 
the essential properties of protein molecules. Proteins have two
local angles per residue (usually denoted $\phi$ and $\psi$), which are
relatively free, and may be considered as the only relevant degrees of freedom.
For almost all amino residues in a polypeptide chain there are two distinct minima of
the local potential energy in the $(\phi, \psi)$ plane, corresponding to the 
$\alpha$ and the $\beta$ local conformation of the chain. The only exceptions 
are glycin which has four minima and proline which has only one. We shall 
use a scalar variable $\phi$ to represent local conformation of the protein.
The local potential energy will be simply modeled by an asymmetric $\phi^4$ 
double well potential, namely
\begin{equation}
  V(\phi) = \varepsilon (\phi + \delta)^2(\phi^2 - {2 \over 3}\phi\delta +
  {1 \over 3}\phi^2 - 2),
\end{equation}
where $\delta$ is the asymmetry parameter, ranging from $-1$ to $1$. The two
minima are positioned at $\phi  = \pm 1$. The energy difference between
the minima is
\begin{equation}
  \Delta E = {16 \over 3} \varepsilon \delta.
\end{equation}
The maxima is positioned at $\phi = -\delta$ and its energy is always zero.
We shall denote by $\theta(x)$ an additional field ($x$ is the position along the
protein), for which solitonic excitations can be realized. Excluding the interaction
with $\phi$, we choose its dynamic to be governed by the Sine-Gordon equation.
It is convenient to picture $\theta$ as an integral over a charge density field.
Indeed, it was suggested that the dynamic of a charge in an ordered linear media
is governed by the SG equation \cite{Charge_Soliton2}, \cite{Charge_soliton_in_DNA1}. 
Nevertheless, many nonlinear
phenomena can by described using equations which allow solitonic excitation, and
the SG dynamics was chosen mainly for practical reasons.

In order that solitons propagating through the protein backbone could mediate
changes in its conformation, an interaction of an appropriate form should be 
introduced. Consider the following interaction potential:
\begin{equation}
  u(\theta,\phi) = {\Lambda \over \Lambda + 1} (1 - \cos\theta)\phi^2,
\end{equation}
where $\Lambda$ is a positive parameter.
This is a natural way to introduce an interaction with the SG model (for example,
interaction with an impurity in a long Josephson transmission line
\cite{Soliton_interaction1}). Note that when $\theta = 0 \bmod 2\pi$ there is no 
interaction. If $\theta$ is non zero ($\bmod 2\pi$), then, at the minima, the energy 
of the combined $V$ plus $u$ potential increases, which effectively lowers the
barrier. We therefore expect that a sufficiently energetic soliton may enable
transition from a meta-stable to a stable conformation (even without thermal noise).

The full Lagrangian density reads:
\begin{equation}
\label{eq_Lagrangian}
  {\cal L} = {1 \over 2} \dot{\theta}^2 - {1 \over 2} \theta_x^2 - 
  {(1 - \cos\theta) \over \Lambda + 1} + {1 \over 2} m  \dot{\phi}^2 - 
  V(\phi) - u(\theta,\phi).
\end{equation}
Note that we used a continues limit for $\phi$.  
This may be a reasonable approximation if the $\theta$ field
changes slowly in space compared with the length of a peptide unit.
To make this model more realistic one may vary the parameters
from point to point along the molecular chain since they depend on the nature
of the local residue. Interaction between different points along the protein,
due to electrostatic and Van der Waals forces, hydrogen bonding etc, can be
introduced by letting the parameters depend on the overall conformation
({\it i.e. \/} they should be functionals of $\phi$). Nevertheless, to test our
proposal for SMF, the simple model will suffice.

From Lagrangian (\ref{eq_Lagrangian}) we get the equations of motion
\begin{eqnarray}
\label{eq_of_motion}
  \ddot{\theta} & = & \theta_{xx} - {1+\Lambda\phi^2 \over 1+\Lambda}\sin \theta - 
                      \Gamma_{\theta}\dot{\theta}  \\
  m\ddot{\phi} & = & -4\varepsilon(\phi+\delta)(\phi^2-1) -
           {2\Lambda \over 1+\Lambda}\phi(1-\cos\theta) - \Gamma_{\phi}\dot{\phi},
\end{eqnarray}
which also include dissipation terms. The scale
was chosen in a way so that when $\phi$ is at a minima ,the coefficient
of the $\sin\theta$ term is exactly $1$. 

First, let us consider the small
interaction limit (we ignore dissipation at this stage). We choose
\begin{eqnarray}
  \theta(x,t) & = & f(\gamma(x-vt)) + \Delta\theta(x,t) \nonumber \\
  \phi(x,t)   & = & 1 + \Delta\phi(x,t),
\end{eqnarray}
where $f(z) = 4\tan^{-1} e^{-z}$ is the usual SG kink soliton moving with velocity
$v$ ($\gamma$ is the relativistic factor). Assuming $\Lambda$ is small and 
expanding to first order one finds:
\begin{equation}
  \Delta\ddot{\phi} + \omega^2 \Delta\phi = 
                   -{4\Lambda\over m}{1\over\cosh^2\gamma(x-vt)},
\end{equation}
with $\omega^2 = 8\varepsilon(1+\delta)/m$. The solution is
\begin{equation}
  \Delta\phi = -{4\Lambda\over\gamma v m\omega} \int_z^{\infty}dz'\, 
               {\sin {\omega\over\gamma v}(z'-z)\over\cosh^2 z'},
\end{equation}
where $z\equiv\gamma(x-vt)$. Asymptotically, for large $z$
\begin{equation}
\label{eq_assimp_dphi}
  \Delta\phi \simeq -{4\Lambda\over m}{e^{-2z}\over(\gamma v)^2 + (\omega/2)^2}.
\end{equation}
For $v=0$ we can write an explicit solution
\begin{equation}
\label{eq_stat_sol}
  \Delta\phi = -{4\Lambda\over m \omega^2 \cosh^2 x}.
\end{equation}
We may compare this to the asymptotic result (\ref{eq_assimp_dphi}). Interpolation
yields the following approximate solitary wave solution:
\begin{equation}
  \Delta\phi \simeq -{4\Lambda/m\over(\gamma v)^2 + (\omega/2)^2}\,
                    {1\over\cosh^2\gamma(x-vt)}.
\end{equation}
As we can see, near the center of the $\theta$ kink, $\phi$ is pushed away from 
its local minima $\phi = 1$ towards the other local minima so we expect that
for strong enough interaction this would allow $\phi$ to cross the energy
barrier.

In order to get further insight into the behavior of this model when interaction
is strong, we set all time derivate terms in (\ref{eq_of_motion}) to zero, 
looking for a static solution. The equations can be solved exactly in the
symmetric ($\delta = 0$) case. we get (note that $\phi=0$ is not a stable
solution)
\begin{equation}
  \phi^2 = 1 -{\Lambda\over2\varepsilon(1+\Lambda)}(1-\cos\theta).
\end{equation}
Inserting this to the equation for $\theta$ we have
\begin{equation}
  \theta_{xx} = (1-a)\sin\theta + a\sin\theta\cos\theta,
\end{equation}
with 
\begin{equation}
  a\equiv {\Lambda^2 \over 2\varepsilon(1+\Lambda)^2}.
\end{equation}
A localized solution is
\begin{equation}
\label{eq_sym_solution}
  \theta  =  4\tan^{-1}{1\over q+\sqrt{1+q^2}} \hspace{1in}
  \phi^2  =  1 - {\Lambda\over\varepsilon(1+\Lambda)}\,{1\over 1+q^2},
\end{equation}
with $q\equiv \sqrt{1-a}\,\sinh(x-x_0)$. We note that no localized solution exist
for $a>1$. This is the strong interaction - low energy barrier regime, where
the solitonic behavior is totally destroyed.
For small asymmetry, we may try to calculate  first order correction
to the above result. Generally, introducing a perturbation $\Delta V$ in the
static symmetric $\phi$ equation, it can be shown that the first order correction
is given by
\begin{equation}
\label{eq_assym_pert_int}
  \Delta\theta(x)  = -{\Lambda \over 4\varepsilon(1+\Lambda)}\int_{-\infty}^x dx'\,
                     G(x-x'){\sin\theta_0(x') \over \phi_0(x')}\Delta V(x'), 
\end{equation}
where $\theta_0$ and $\phi_0$ are the solutions (\ref{eq_sym_solution}) with
$x_0=0$, and the Green function is
\begin{equation}
  G(x) = {1\over2}\left((1-a)\sinh x + {(3a-1)a\sinh x + (3a+1)(1-a)x\cosh x
         \over 1 + (1-a)\sinh^2 x}\right).
\end{equation}
unfortunately, for most of the interesting perturbations, including the
asymmetric barrier perturbation
\begin{equation}
  \Delta V = 4\varepsilon\delta(\phi_0^2-1), 
\end{equation}
the integral (\ref{eq_assym_pert_int}) can not be calculated analytically.

\begin{figure}
  \centerline{\psfig{figure=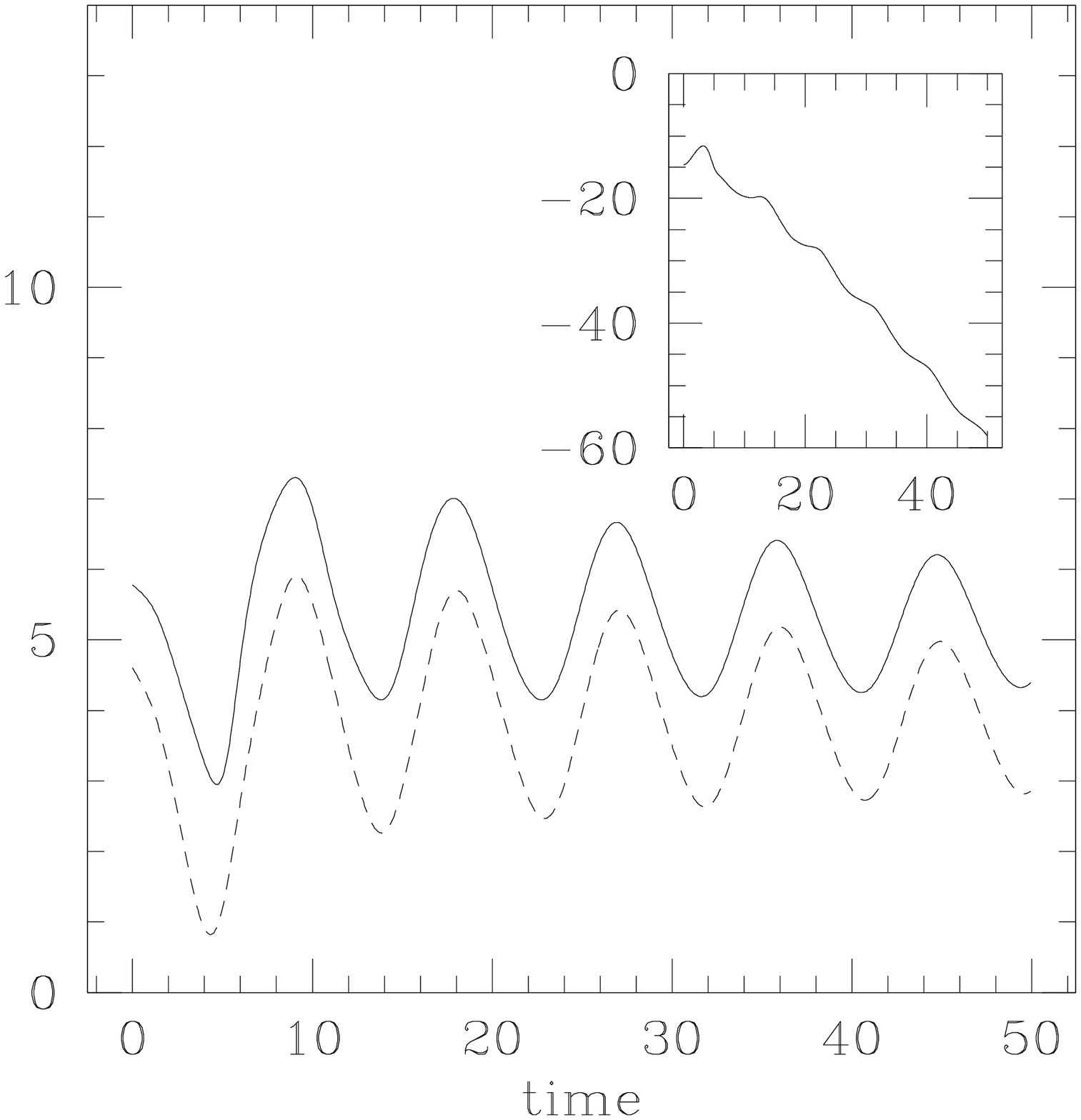,height=6.5cm}}
  \vspace{0.25in}
  \caption{The collective momentum and the energy of a soliton moving through a
  protein which is initialy in a metastable (unfolded) state. The momentum is indicated
  by the dashed line. Continues line
  indicates soliton energy. The energy of the conformation field is shown in the small
  box. The parameters have the following values: $\varepsilon=1,m=10,\Lambda=100,
  \delta=-0.5, \gamma_{\phi} = 0.1, \Gamma_{\theta} = 0.1$. Note that, although
  dissipation is introduced, the soliton momentum has a constant positive average value.}
\label{P_and_E_fig1}
\end{figure}

We shall turn now to discuss results of numerical simulations for the model.
When a topological constraint was enforced upon the $\theta$ field (a difference
of $2\pi$ between the boundaries) and the system was allowed to relax via 
dissipation to a static conformation, a result similar to that of equation
(\ref{eq_stat_sol}) was observed. Next we considered initial conditions of a 
moving $\theta$ kink, with $\phi=1$ (local minima) for all x. Though it seems
reasonable, based on the form of the interaction potential, that a traveling
soliton might transfer enough energy to the $\phi$ field to lift it over the
energy barrier, it is certainly not obvious a priori whether this energy would be 
returned to the soliton. Note that interacting soliton also radiates plasmons, so 
it losses energy through this channel too (even without dissipation).
It is useful to define collective coordinate and momentum for a soliton
\begin{equation}
  Q = -{1\over2\pi}\int_{-\infty}^{\infty}dx\,x\theta_x \hspace{1in}
  P = -\int_{-\infty}^{\infty}dx\,\theta_x\dot{\theta}.
\end{equation}
These are in fact canonical variables which represent the soliton's `center of
mass' and its conjugate momentum. The relation $\dot{Q}\simeq P/E$ holds, with $E$
being the total energy of the soliton (deviations from the exact formula are due
to deformation of the interacting soliton shape). We used $P$ as an indicator for
the soliton's kinetic energy.  Simulations reveal that when $\delta=0$ 
({\it i.e.\/} symmetric barrier) the kink slows down and stops after a short period of
time. Next, $\delta$ is decreased so that the $\phi=-1$ conformation
becomes lower in energy. When interaction is strong enough, the kink transfers enough
energy to the conformation angles, allowing them to cross the potential barrier and reach
the ground state ($\phi=-1$), in practically every point it passes. However, the
soliton extract back some of the conformation energy. After short period of time it
reaches a steady state finite velocity (up to some periodic fluctuations), when the
energy gained balances the energy lost. When dissipation
is introduces, it lowers the steady state velocity, but the overall picture
remains the same. Results of simulation are shown in fig. \ref{P_and_E_fig1}. We
conclude that within this model, a soliton is able to mediate
folding ({\it i.e. \/} carry conformational degrees of freedom over energy barriers)
and in the process compensate for energy losses by pumping it back from the
folded field.

\begin{figure}
  \centerline{\psfig{figure=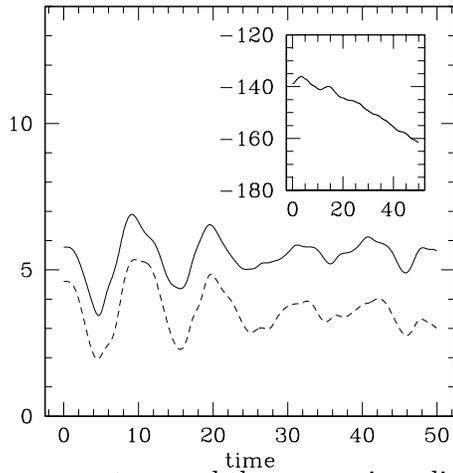,height=6.5cm}}
  \caption{The collective momentum and
  the energy in a disordered initial conditions. Dissipation parameter $\Gamma_{\theta}$
  was set to zero. All other parameters have the same value as in
  Fig. {\protect \ref{P_and_E_fig1}} . The protein conformation energy is shown is
  the small box.}
\label{P_and_E_fig2}
\end{figure}

In actual unfolded state local folding angles are supposed to be
distributed somehow between $\alpha$ and $\beta$ conformation. We have checked
the behavior of a traveling kink in a disordered initial conditions, namely,
when the $\phi$ field is randomly distributed between the two minima. Still,
a steady state velocity is reached, though lower compared with the ordered case
(see fig. \ref{P_and_E_fig2}).
This can be easily explained since less compensating energy is available to the
soliton when already half the conformational degrees of freedom are at their
ground state. Nevertheless, the folding mediated is not less effective. For
sufficient asymmetry all the angles reach the lower energy. The
soliton did not only lowered the energy but also the entropy.

We have demonstrated, using a simple toy model, how soliton, propagating along
a protein backbone provides an effective mechanism for fast and deterministic 
folding of the
protein. By mediating correlated fold of large sections of the polypeptide
chain, it reduces significantly the number of effective degrees of freedom, since
large structures may be created by a single soliton. Moreover, these structures
are created in a controlled process. It is possible that the amino 
acids sequence dictates whether soliton propagation would be allowed or suppressed
in a specific location at a specific time. therefore the folding pathway may not
be stochastic but, rather, predetermined. The limiting rate of the
folding process would be the rate in which solitons are produced. Soliton creation
is probably induced spontaneously by thermal
excitation. There might exist other mechanisms for creation of solitons, such as 
interaction with `chaperones', enzymes that catalyze protein folding. It is our belief
that the concept of SMF will improve scientific understanding of such phenomena.

The authors wish to thank Ziv~Hermon for valuable discussions. This work has been
partly supported by GIF grant.

\end{document}